\documentclass[%
 aip,
 amsmath,amssymb,
 reprint,%
]{revtex4-1}

\usepackage{graphicx}
\usepackage{dcolumn}
\usepackage{bm}

\usepackage[normalem]{ulem}
\usepackage[utf8]{inputenc}
\usepackage[T1]{fontenc}
\usepackage{mathptmx}
\usepackage{etoolbox}
\usepackage{xcolor}
\usepackage{comment}

\newcommand{\UP}{\pmb{\Uparrow}}
\newcommand{\DOWN}{\pmb{\Downarrow}}

\makeatletter
\def\@email#1#2{%
 \endgroup
 \patchcmd{\titleblock@produce}
  {\frontmatter@RRAPformat}
  {\frontmatter@RRAPformat{\produce@RRAP{*#1\href{mailto:#2}{#2}}}\frontmatter@RRAPformat}
  {}{}
}%
\makeatother
\begin{document}

\preprint{AIP/123-QED}

\title{Second-order Raman and anharmonic lattice dynamics in lead-halide perovskites}
\author{Chao Shen}

\affiliation{ 
Institute of Science and Technology Austria, Am Campus 1, Klosterneuburg 3400, Austria
}%
\author{Florian Kluibenschedl}

\affiliation{ 
Institute of Science and Technology Austria, Am Campus 1, Klosterneuburg 3400, Austria
}%

\author{Ayan A. Zhumekenov}
\affiliation{%
Department of Chemistry, King Fahd University of Petroleum and Minerals, Dhahran 31261, Saudi Arabia
}
\affiliation{%
Interdisciplinary Research Center (IRC) for Advanced Materials, King Fahd University of Petroleum and Minerals, Dhahran 31261, Saudi Arabia
}
\author{Mikhail Lemeshko}

\affiliation{ 
Institute of Science and Technology Austria, Am Campus 1, Klosterneuburg 3400, Austria
}%

\author{Osman M. Bakr}
\affiliation{%
King Abdullah University of Science and Technology, Thuwal, Kingdom of Saudi Arabia
}
\affiliation{%
Center for Renewable Energy and Storage Technologies,
Division of Physical Science and Engineering,
King Abdullah University of Science and Technology, Thuwal, Kingdom of Saudi Arabia
}
 
 \author{Artem G. Volosniev}
\affiliation{%
Department of Physics and Astronomy, Aarhus University, Ny Munkegade 120, Aarhus DK-8000, Denmark
}%
\author{Zhanybek Alpichshev}%
 \email{alpishev@ist.ac.at}
\affiliation{ 
Institute of Science and Technology Austria, Am Campus 1, Klosterneuburg 3400, Austria
}%

\date{\today}

\begin{abstract}
We explore the microscopic origin of THz-driven lattice dynamics in methylammonium lead bromide CH$_3$NH$_3$PbBr$_3$ using two-dimensional THz Kerr-effect spectroscopy. The analysis of the experiment enables discrimination between competing nonlinear excitation mechanisms and reveals that the dominant pathway is phonon-mediated rather than purely photonic. In particular, we argue that the commonly observed 1.15~THz Raman-active $A_g$ mode is not excited directly by the THz electric field. Instead, it emerges through cubic anharmonic coupling to an infrared-active phonon at 0.68~THz. Furthermore, a satellite feature at 1.36~THz is identified as a two-phonon Raman response of the same infrared-active mode. The observation of strong quadratic coupling involving the infrared-active phonon suggests an unconventional form of electron-phonon dressing present in lead-halide perovskites. 
\end{abstract}

\maketitle

\section{Introduction}
Lead halide perovskites exhibit remarkable optoelectronic properties, including efficient photoluminescence~\cite{Kovalenko2017,Chouhan2020}, tolerance to structural disorder~\cite{MosqueraLois2025}, and, perhaps most strikingly, a phenomenology consistent with collective quantum-coherent behaviour persisting up to room temperature~\cite{Biliroglu2025}. Together with other experimental observations~\cite{Kim2017,Guo2020,Sendner2016,Zhu2016,Miyata2017}, many of these findings point to a central role of lattice dynamics in these materials and motivate mode-resolved investigations of the coupling between electronic and lattice degrees of freedom. 

In this Letter, we explore the electron-phonon coupling in the hybrid metal-organic lead halide perovskite $\mathrm{CH_3NH_3PbBr_3}$   ($\mathrm{MAPbBr_3}$) using an ultrafast probe in which intense terahertz (THz) pulses excite coherent lattice vibrations that modify the electronic polarizability. These modifications are detected using a near-infrared probe pulse that measures THz-pump-induced changes in the sample's refractive index via the Kerr effect.

In order to efficiently observe electron-phonon coupling, one has to be able to drive the lattice degrees of freedom strongly. This, however, inevitably leads to parasitic anharmonic coupling between different lattice modes. To overcome this, we employ two THz pump pulses with a controllable delay between them, i.e., two-dimensional THz Kerr-effect (2D-TKE) spectroscopy. Combined with a coupled-oscillator model, this approach allows us to resolve correlations between excitations and distinguish different nonlinear excitation pathways. In particular, we disentangle the contributions of a Raman-active mode at $1.15~\mathrm{THz}$, associated with twisting of the lead-bromide octahedra, and a previously unreported feature at $1.36~\mathrm{THz}$. Our analysis reveals that the latter is not driven directly by the THz field but instead emerges through cubic anharmonic coupling with an IR-active mode at $0.68~\mathrm{THz}$, which generates a two-phonon Raman response at this frequency. These findings demonstrate the central role of anharmonic multiphonon processes in the ultrafast optical response of hybrid perovskites.

The observation of strong second-order coupling between IR-active phonons and electronic polarizability may play an important role in the dressing of photoexcitations, including charge-neutral electron-hole pairs (excitons). In the Outlook, we discuss a possible dressing mechanism arising from this interaction, which may have direct implications for the interpretation of recent observations of quantum protection of radiative dipoles in lead-halide perovskites~\cite{Biliroglu2025}. A detailed theoretical and experimental investigation of this mechanism is left for future work.

\section{Experiment}

\begin{figure}[t!]
\includegraphics[scale=0.6]{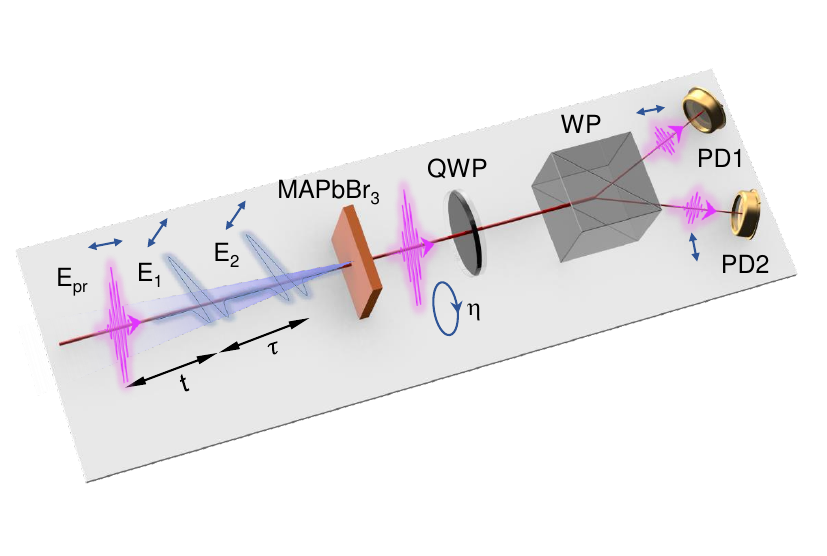}
\caption{\label{fig:1} Schematic illustration of the 2D TKE experiment. QWP: quarter-wave plate; WP: Wollaston prism; PD: photodiode.}
\end{figure}

The experimental setup of 2D-TKE is illustrated in Fig.~\ref{fig:1}. Two intense THz pulses ($E_\mathrm{1}$ and $E_\mathrm{2}$) were generated by pumping two 70 fs ultrashort pulses at 800 nm on two separate LiNbO$_3$ crystals using the tilted-pulse front method. A pair of wire-grid polarizers were used to combine the two THz pulses and modify the polarization orientations of THz electric fields. The MAPbBr$_3$ single crystals used for the study were grown by inverse temperature crystallization method, as described elsewhere~\cite{saidaminov2015high}. The sample was mounted in a helium flow cryostat with one of the sample axes oriented at \(45^\circ\) to the THz electric fields. The peak field strengths of THz 1 and THz 2 pulses were estimated to be $\sim300$ kV/cm and $\sim450$ kV/cm at the focus, respectively. THz-induced Kerr effect was measured in the transmission geometry and the time-resolved ellipticity signal of an 800 nm probe pulse, $E_\mathrm{pr}$, was measured by a combination of a quarter-wave plate, a Wollaston prism and a balanced detector. The polarization of the probe pulse was parallel to the sample axis. A differential chopping scheme \cite{lu2017coherent} was employed to extract the nonlinear signal $\eta_\mathrm{NL} = \eta_\mathrm{12} - \eta_\mathrm{1} - \eta_\mathrm{2}$, where $\eta_\mathrm{12}$ and $\eta_{i}$ are the ellipticity signals measured with both THz pulses present and with only pulse \(i\) present, respectively. In single pulse TKE measurements, only THz 2 is used (the THz electric field is aligned with the sample axis) and the signal is acquired using a standard pump-probe scheme. Further details of the experiment can be found elsewhere \cite{shen2025nonlinear,shen2026disentangling}. 

\section{Results}

\begin{figure}[t!]
\includegraphics[scale=0.43]{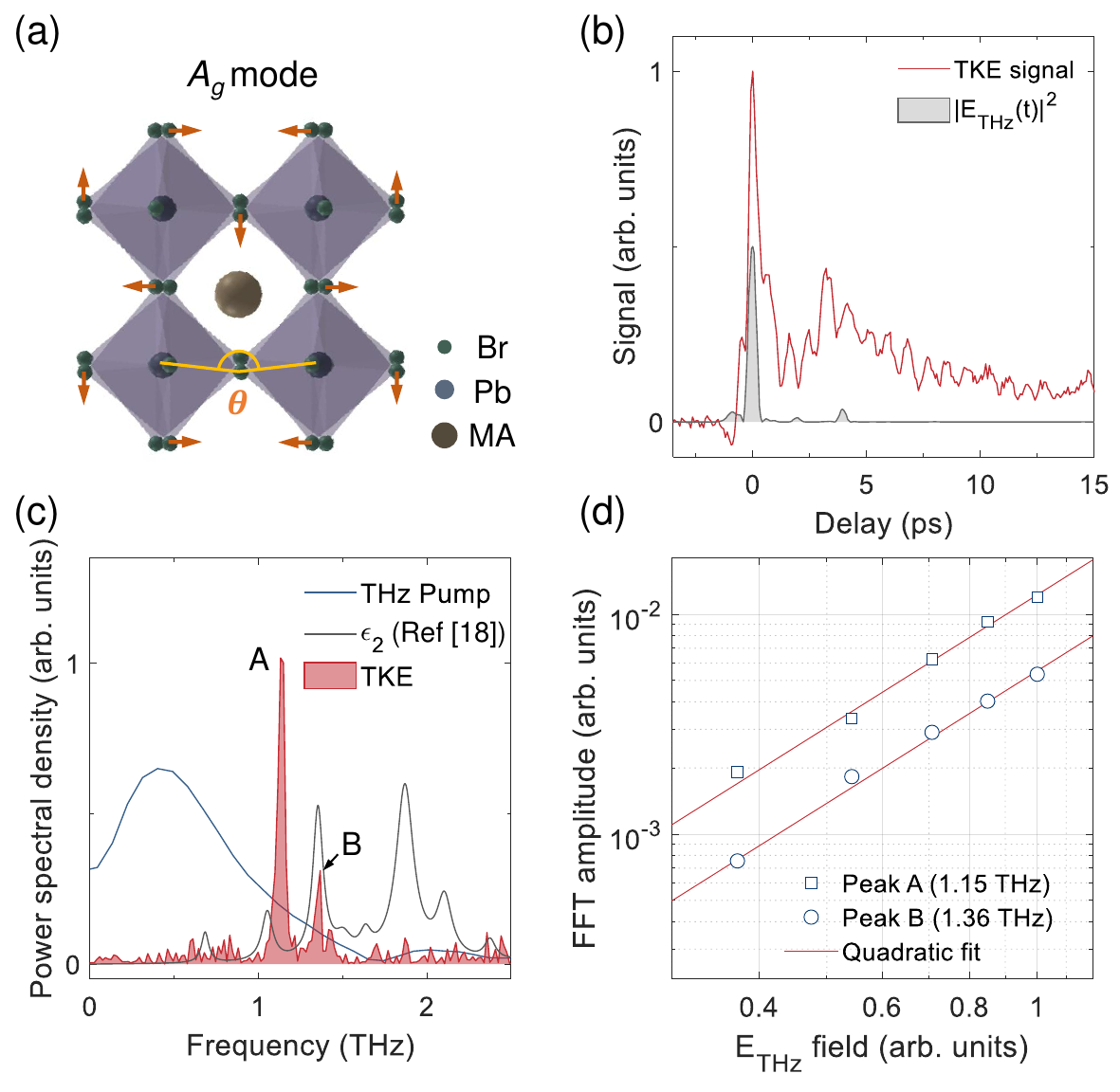}
\caption{\label{fig:2} A$_g$ mode oscillation and single pulse TKE measurement of MAPbBr$_3$ at $T=80$~K. (a) A$_g$ mode which is associated with the modulation of the Pb-Br-Pb bond angle $\theta$. (b) Time domain TKE response, together with the squared THz electric field $|E_{\text{THz}}(t)|^2$ measured by the electro-optic sampling using a 200 $\mu$m-thick ZnTe. (c) Power spectral density of TKE response together with THz spectral amplitude and the imaginary dielectric constant $\varepsilon_2$ of MAPbBr$_3$ (adapted from Ref.\cite{vzelezny2023infrared}). To eliminate the contribution from the electronic hyperpolarizability, the differentiated time domain signal is used. (d) THz electric field and quadratic fit of peaks A and B mode amplitude. }
\end{figure}

We first revisit the single pulse TKE result at $T=80$~K, which is shown in Fig.~\ref{fig:2}. In the time-domain trace (Fig.~\ref{fig:2}(b)), the signal starts with a rapidly decaying peak at time zero followed by an oscillatory signal lasting tens of picoseconds. The peak exists only when the THz pulse is present (see the comparison with $E^2_{\text{THz}}(t)$), indicating an origin of instantaneous electronic response. The fast Fourier transform (FFT) analysis of the oscillating signal in Fig.~\ref{fig:2}(c) shows two pronounced frequency components, i.e., peak A at $\sim$1.15~THz and peak B at $\sim$1.36~THz. Peak A has been extensively observed and analyzed in hybrid lead perovskites and has been attributed to the Raman-active octahedral twisting mode with $A_g$ symmetry \cite{leguy2016dynamic,guo2020direct,frenzel2023nonlinear} (Fig.~\ref{fig:2}(a)). Peak B so far has not been discussed although it has been observed in previous TKE measurements \cite{frenzel2023nonlinear} and also in a static low-temperature Raman scattering experiment \cite{yaffe2017local}. The nonlinear origin of both modes is substantiated by the quadratic scaling with THz electric field strength, as shown in Fig.~\ref{fig:2}(d). By inspecting the THz spectra and the imaginary part of the dielectric function of MAPbBr$_3$ at low temperature, as shown in Fig.~\ref{fig:2}(c), we see that our THz pulse has a center frequency at 0.5~THz with a bandwidth of $\sim$0.6~THz and it covers several IR-active phonons. Thus, both THz sum frequency excitation and IR phonon-assisted coupling are possible.  

\begin{figure}[t!]
\includegraphics[scale=0.5]{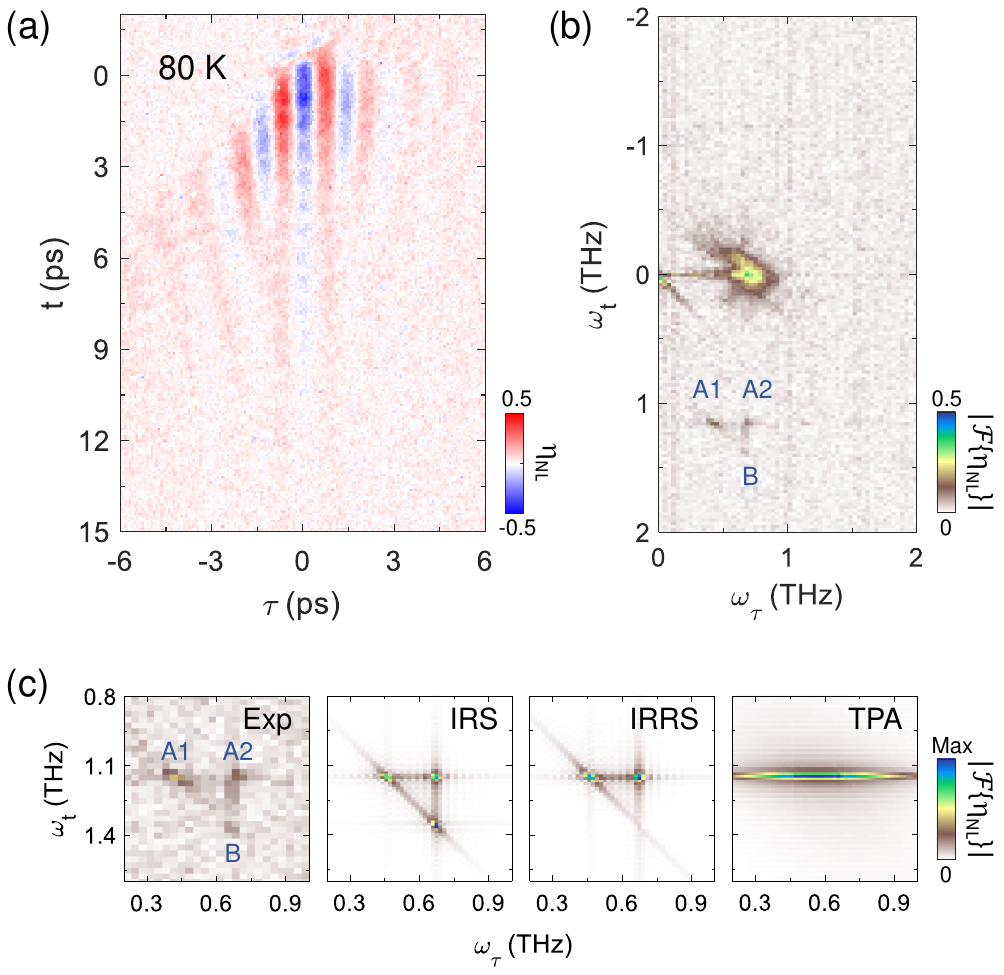}
\caption{2D-TKE measurements and simulations. (a,b) Experimental 2D-TKE data at $T=80~\mathrm{K}$ in the time and frequency domains, respectively. (c) Enlarged view of the experimental three-peak structure and corresponding simulations using the driving terms associated with IRS, IRRS, and TPA.}
\label{fig:3}
\end{figure}

To reveal the origin of the observed dynamics, we performed the 2D-TKE measurements at the same temperature. The 2D Kerr  response in time domain shows a strong oscillation along $\tau$ axis and a weaker but still discernible oscillation along $t$ axis, as shown in Fig.~\ref{fig:3}(a). A 2D FFT analysis shows that the stronger oscillation corresponds to a peak at ($\sim$0.68, 0) THz in the $(\omega_\tau,\omega_t)$ plane. This peak corresponds to a THz rectification process, in which a coherent oscillation at 0.68 THz is generated by the first THz pulse and quenched by the second one. More importantly, we observe a unique three-peak structure, as marked by A1 ($\sim$0.45, $\sim$1.15) THz, A2 ($\sim$0.68, $\sim$1.15) THz and B ($\sim$0.68, $\sim$1.36) THz. These features further indicate that a genuine oscillation at 0.68 THz is excited by THz pulses. Interestingly, this frequency excellently matches that of an IR phonon as reported by a recent THz reflectivity measurement \cite{vzelezny2023infrared}, strongly hinting at  phonon-phonon coupling as the origin of the observed dynamics.

\section{Discussion}

In centrosymmetric materials, Raman-active phonons carry zero dipole moment and can only be excited through higher-order processes. Underlying excitation pathways include two-photon absorption (TPA) \cite{maehrlein2017terahertz}, ionic Raman scattering (IRS) \cite{forst2011nonlinear,juraschek2018sum,melnikov2018coherent,li2019terahertz,henstridge2022nonlocal} and the recently proposed hybrid photo-phononic process, also dubbed infrared-resonant Raman scattering (IRRS) \cite{khalsa2021ultrafast}. As these mechanisms often obey the same symmetry constraints and exhibit similar quadratic scaling with THz electric field, simple analysis based on polarization selection rules or field dependence is generally insufficient to distinguish the dominant pathway\cite{johnson2019distinguishing,knighton2019terahertz}.

We identify the specific excitation pathway of the Raman response by comparing \textcolor{black}{the} 2D-TKE experiments with theoretical modeling. At $T=80$~K, MAPbBr$_3$ is in the orthorhombic phase with a point group $D_{2h}$. Any nonlinear coupling terms through the lattice potential should therefore transform according to the fully symmetric representation of the $D_{2h}$ point group. 

We adopt the following minimal symmetry-allowed nonlinear lattice potential up to cubic order in the coupling terms to model the nonlinear Kerr response:
\begin{equation}
    \begin{split}
         U = & - Z^* Q_{\text{IR}}E - a Q_{\text{IR}}^2 Q_{\text{R}} - b Q_{\text{IR}} Q_\text{R} E - cQ_\text{R}E^2,
    \end{split}
    \label{eq1}
\end{equation}
where $Z^*$ is the Born effective charge of the IR mode and $a$, $b$ and $c$ are phenomenological constants. The last three terms on the right-hand side of the Eq.~(\ref{eq1}) indicate the IRS, IRRS and TPA mechanisms, respectively. Taking the derivative of $U$ with respect to $Q_\text{IR}$ and $Q_\text{R}$, we obtain the following coupled equations of motion for the Raman and IR modes \\
\begin{align}
    \mathcal{L}_\text{IR}Q_\text{IR}  &= Z^*E + 2aQ_\text{IR}Q_\text{R} + bQ_\text{R}E \label{eq:equation_of_motion_IR_mode}, \\
    \mathcal{L}_\text{R}Q_\text{R}  &= aQ^2_\text{IR} + bQ_\text{IR}E + cE^2
    \label{eq:equation_of_motion_Raman_mode},
\end{align}
where we have introduced the linear differential operator ${\mathcal{L}}_\alpha = M_\alpha[(d^2/dt^2)+2\gamma_\alpha\omega_\alpha(d/dt)+\omega_\alpha^2]$; $M_\alpha$, $\gamma_\alpha$ and $\omega_\alpha$ are the reduced mass, normalized damping constant and frequency of the mode $\alpha$, respectively. The nonlinear Kerr signal is given by: 
\begin{equation}
    \eta_\text{NL} \propto Q_{\text{R},\text{12}}^{} - Q_{\text{R},\text{1}}^{} - Q_{\text{R},\text{2}}^{},
\end{equation}
where $Q_{\text{R},\text{1}}^{}$, $Q_{\text{R},\text{2}}^{}$ and $Q_{\text{R},\text{12}}^{}$ stand for the $A_g$ phonon coordinate excited by THz 1 or THz 2 pulse alone, or with both together, respectively. 

\indent Figure~\ref{fig:3}(c) shows simulated nonlinear Kerr responses in the frequency domain using three driving terms respectively. In the simulation, we use $\omega_{\text{R}}/2\pi=1.15$~THz, $\omega_{\text{IR}}/2\pi=0.68$~THz. We set the normalized Raman-mode damping constant to $\gamma_\mathrm{R} = 0.005$ to match the phonon lifetime extracted in the single pulse measurements. The damping of the IR mode at $T = 80~\,\mathrm{K}$ has not been reported; we therefore adopt $\gamma_\mathrm{IR} = 0.006$, as measured at $T = 10~\,\mathrm{K}$~\cite{vzelezny2023infrared}. We emphasize that the results presented below remain valid under variations of $\gamma_\mathrm{IR}$ within an order of magnitude (see Appendix \ref{sec:ir_gamma_dep}). In all simulations, we use a THz waveform fitted from the electro-optic sampling measurement on a \(200~\mu\mathrm m\)-thick, (110)-cut ZnTe crystal. The remaining parameters $M_\alpha$ and $Z^*$ can be chosen arbitrarily, as their absolute values do not influence the peak positions in the simulated TKE response. In the following, both quantities are set to unity in their respective units. 

The purely photonic TPA mechanism can be readily excluded, as it produces a continuous, line-like feature along $\omega_t = \omega_{\mathrm{R}}$, rather than discrete peaks. This behavior arises from the large bandwidth of the excitation THz pulse. As long as the energies of the participating photon pair satisfy energy conservation, the process is allowed, resulting in a continuous distribution in the frequency domain. The spectral weight of the pump pulse then determines the intensity distribution along this line~\cite{mead2020sum}. While the combined photo-phononic IRRS mechanism reproduces peaks A1 and A2, it fails to account for peak B. In contrast, the purely phononic IRS, despite being off-resonant, successfully captures all three peaks. The fact that both IRS and IRRS reproduce peaks A1 and A2 can be understood as follows: in the non-resonant regime, the IR-active mode undergoes a forced oscillation driven by the external field, such that the phononic driving term $\propto Q_{\rm IR}^2$ effectively reduces to a photo-phononic form $\propto Q_{\rm IR} E$. Consequently, peaks A1 and A2 obtained within the IRS framework share the same physical origin as those in the IRRS mechanism \cite{blank2023two}.

To gain further insight into the numerical results, we solve the equation of motion for the Raman coordinate [Eq.~\eqref{eq:equation_of_motion_Raman_mode}] for different driving terms. Neglecting the backaction of the Raman mode on the IR mode, the IR coordinate freely oscillates at its eigenfrequency $\omega_{\rm IR}$ following excitation by a short pulse. For two pulses separated by a delay $\tau$, this yields
$
Q_{\mathrm{IR}}(t,\tau)
\varpropto
\sin(\omega_{\mathrm{IR}} t)
+
\sin(\omega_{\mathrm{IR}} (t+\tau))
$. 
The resulting nonlinear driving, $\propto Q_{\rm IR}^2$, contains components oscillating as $e^{\pm i \omega_{\rm IR}\tau}$ and $e^{\pm i (2\omega_{\rm IR} t + \omega_{\rm IR}\tau)}$. Solving the driven harmonic oscillator for these contributions yields the nonlinear Raman response. In particular, oscillations at $(\omega_{\tau}, \omega_t) = (\pm \omega_{\rm IR}, \omega_{\rm R})$ and $(\omega_{\tau}, \omega_t) = (\omega_{\rm IR}, 2\omega_{\rm IR})$ correspond to peaks A2 and B in Fig.~\ref{fig:3} (see App.~\ref{subsec:raman_after_two_pulse_excitation} for details). Peak B can thus be identified as a two-phonon Raman response of the 0.68~THz mode. In contrast, immediately after excitation, the IR mode undergoes a forced oscillation, $Q_{\rm IR}(t) \propto E(t)$, such that the off-resonant driving term $\propto Q_{\rm IR}^2$ effectively reduces to $Q_{\rm IR} E$. Owing to the broad bandwidth of the THz pulse, this mixed term can satisfy the resonance condition of the Raman mode, yielding a response at $(\omega_{\tau}, \omega_t) = (\omega_{\rm R} - \omega_{\rm IR}, \omega_{\rm R})$, corresponding to peak A1 in Fig.~\ref{fig:3} (see App.~\ref{subsec:raman_mode_after_field}).

\section{Summary and Outlook}

\subsection{Summary}
Using 2D-TKE spectroscopy, we identified the microscopic origin of THz-driven Raman dynamics in MAPbBr$_3$. We argued that the commonly observed Raman mode at 1.15~THz is not excited directly through two-photon THz absorption but is instead driven via anharmonic mixing with an infrared-active mode at 0.68~THz. Furthermore, the feature at 1.36~THz was shown to be the two-phonon Raman response mediated by the same infrared-active mode. 

The characteristic structure observed in the 2D TKE spectra provides a direct fingerprint of phonon-mediated nonlinear excitation pathways and enables their distinction from competing photonic mechanisms. These findings establish 2D THz spectroscopy as a powerful approach for disentangling nonlinear dynamics in complex materials with anharmonic lattices.

\subsection{Outlook: non-Fr\"ohlich coupling and Quantum Polarons}

\begin{figure}[t!]
\includegraphics[width=\columnwidth]{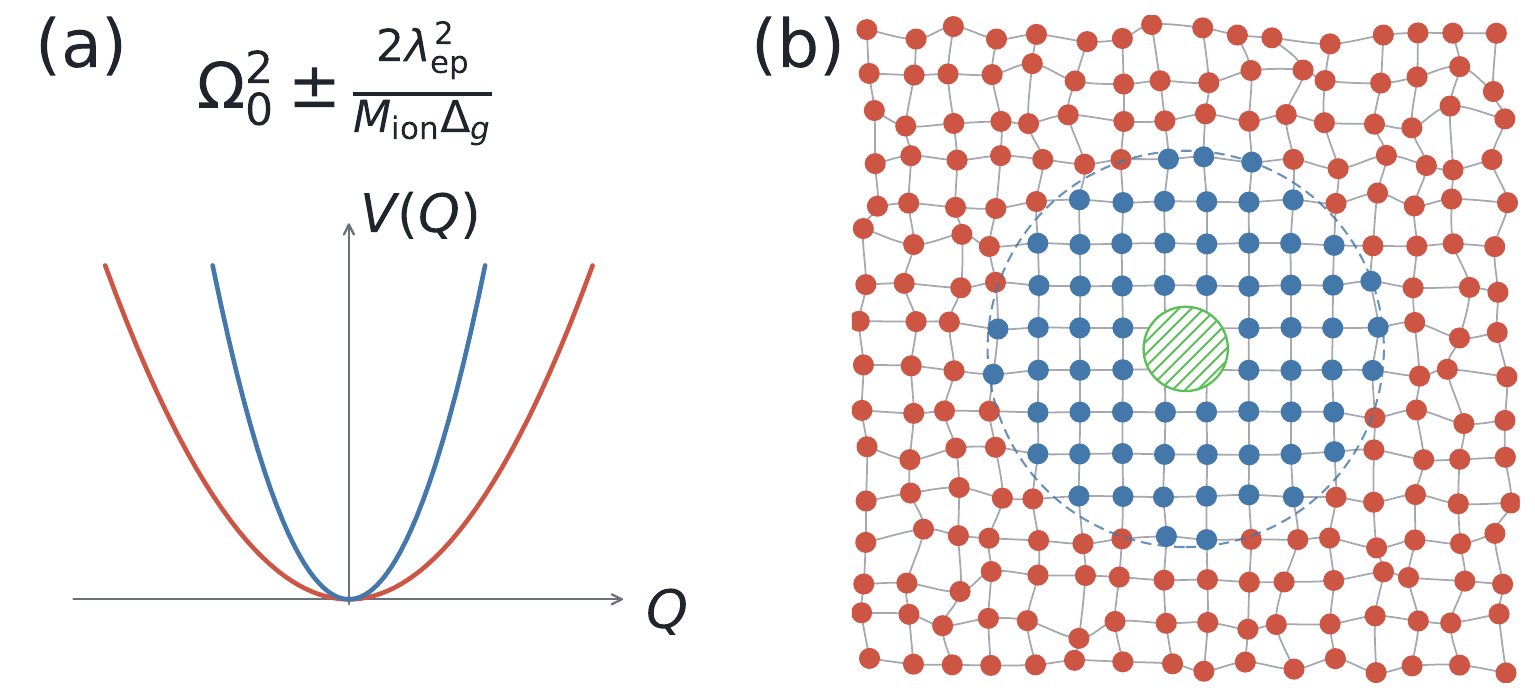} 
\caption{Quantum polaron: An electron suppresses local lattice fluctuations that results in renormalization of the phonon frequency (panel (a)). This renormalization gives rise to a more ordered lattice environment in the vicinity of an electron (red circle in panel (b)). The resulting phonon ordering may form a protective layer that enhances the coherence of quasiparticles in lead-halide perovskites, including excitons.}
\label{fig:quantum_polaron}
\end{figure}

The results of this work suggest the presence of a strong effective electron-phonon coupling $\propto Q_{\mathrm{IR}}^2$, distinct from the standard Fr{\"o}hlich interaction, which is linear in the phonon field~\cite{Devreese2009}. This interaction gives rise to an unconventional dressing of electronic degrees of freedom. In particular, it has been proposed that a strong quadratic electron-phonon interaction can support a novel form of quasiparticle dressing, termed the {\it quantum polaron}~\cite{Gogolin1991, Gogolin1991a}. Unlike conventional polarons, whose formation is accompanied by lattice polarization around the charge carrier, the quantum polaron is characterized by the suppression of local quantum lattice fluctuations (see Fig.~\ref{fig:quantum_polaron}). 

To illustrate why lead-halide perovskites may provide a favorable platform for realizing quantum polarons, we briefly examine the microscopic structure of the electron--lattice interaction using the phenomenological symmetry-based framework developed in Refs.~\cite{Volosniev2023,Volosniev2023a}. In particular, for the high-temperature cubic perovskite structure, we propose that symmetry imposes strong constraints on the allowed coupling between band electrons and polar lattice distortions $\vec{Q}$, leading to the non-Fr\"ohlich interaction \begin{equation} H_{\rm el-ph} = \lambda_{\rm ep}\,\tau_1 \otimes \vec{\sigma}\cdot \vec{Q}, \label{eq:hamiltonian_spin_electric} \end{equation} where Pauli matrices $\tau_\alpha$ and $\vec{\sigma}=\{\sigma_\beta\}$ describe the orbital and spin degrees of freedom of the band electrons, respectively, as discussed in Refs.~\cite{Volosniev2023,Volosniev2023a}; $\lambda_{\rm ep}$ characterizes the strength of the electron--phonon interaction. A key feature of the coupling in Eq.~(\ref{eq:hamiltonian_spin_electric}) distinguishing it from conventional Fr\"ohlich term is that, although it is also linear in $\vec{Q}$, it mixes electron and hole sectors.

To determine the effect of the electron-phonon interaction in Eq.~(\ref{eq:hamiltonian_spin_electric}) within the conduction-band (or valence-band) manifold alone, one can treat $H_{\rm el-ph} $ perturbatively, which generates the correction to the effective Hamiltonian,


\begin{equation}
\delta H = \frac{\lambda_{\rm ep}^{2}Q_{IR}^{2}}{\Delta_g} \tau_3 \otimes \sigma_0,
\label{eq:deltaH}
\end{equation} 

\noindent where $\Delta_g$ is the electronic band gap, see App.~\ref{sec:foldy_wouthuysen_transformation} for derivation details. Since this interaction is quadratic in the phonon coordinate, it is manifestly non-Fr\"ohlich and therefore does not induce a static lattice polarization around a charge carrier.

As shown by Gogolin and Ioselevich~\cite{Gogolin1991,Gogolin1991a}, the quadratic character of lattice coupling in Eq.~(\ref{eq:deltaH}) has qualitatively distinct effects as compared to Fr\"ohlich-type linear coupling, causing reconstruction of the lattice environment around electronic excitation characterized by the suppression of local quantum fluctuations~\cite{hanQuantumBipolaronSuperconductivity2024a}. The resulting quasiparticle, quantum polaron, differs qualitatively from the conventional Landau-Pekar polaron~\cite{Landau}. In particular, quantum polarons do not involve lattice polarization around its core, and can  therefore form around charged (electrons, holes) and {\it neutral} excitations alike (excitons). Moreover, if the lattice stiffening within the quantum polaron is sufficiently strong, it may suppress thermal lattice fluctuations near the polaron core. Such a mechanism could provide a natural explanation for the apparent protection of coherent optical dipoles in lead-halide perovskites at elevated temperatures. 

Establishing the relevance of the quantum-polaron scenario for lead-halide perovskites in real life requires a quantitative determination of the interaction strength $\lambda_{\rm ep}$, which we leave to future studies. However, we can already estimate the potential magnitude of lattice stiffness renormalization caused by electronic excitation based on general considerations. In particular, using $H_{\rm el-ph}$ above, the upper bound on the phonon-frequency renormalization can be written as:

\begin{equation} \Omega_{\pm}^{2} = \Omega_{0}^{2} \pm \frac{2\lambda_{\rm ep}^{2}}{M_{\rm ion}\Delta_g}, \end{equation} 

\noindent where $M_{\rm ion}$ is the effective ionic mass associated with the relevant phonon mode, $\Omega_0$ is the bare phonon frequency, and the signs $-$ and $+$ correspond to the electronic ground and excited states, respectively. To estimate the magnitude of renormalization effect, we use: $\lambda_{\rm ep}\sim \Delta_g/\tilde a$, where $\tilde a$ is the lattice constant. Taking $M_{\rm ion}\simeq M_{Br}$, we obtain: 

\begin{equation} \sqrt{\frac{2\lambda_{\rm ep}^{2}} {M_{\rm ion}\Delta_g}} \approx 5\,\mathrm{THz}. \end{equation} 

\noindent Taking, for definiteness, a phonon frequency at equilibrium $\Omega_- \lesssim 3\,\mathrm{THz}$, consistent with experimental observations, we estimate a renormalized frequency inside the quantum-polaron cloud to be as high as $\Omega_+ \sim 8\,\mathrm{THz}$. Such a large shift should be readily detectable in dedicated experiments, for example by time-domain THz spectroscopy following optical excitation of electron--hole pairs. The substantial ratio $\Omega_+/\Omega_-$ predicted for lead-halide perovskites could account for both the anomalous softness of the lattice at equilibrium and the apparent protection of coherence of optical dipoles in these materials.

\begin{acknowledgments}
We thank Sebastian F. Mährlein, Ragheed Al Hyder, and Simon Wall for stimulating discussions. A.~G.~V. acknowledges support from the Novo Nordisk Foundation
(grant reference number NNF25OC0102659).  Z.A. acknowledges support from the collaborative research project SFB Q-M\&S funded by the Austrian Science Fund (FWF, grant No.~PR1050F8602). 
\end{acknowledgments}

\section*{Data Availability Statement}

The data that support the findings of this study are available from the corresponding author upon reasonable request.

\appendix

\section{The role of damping of IR-active phonon} \label{sec:ir_gamma_dep}

\begin{figure}[htbp]
    \centering
    \includegraphics[width=\columnwidth]{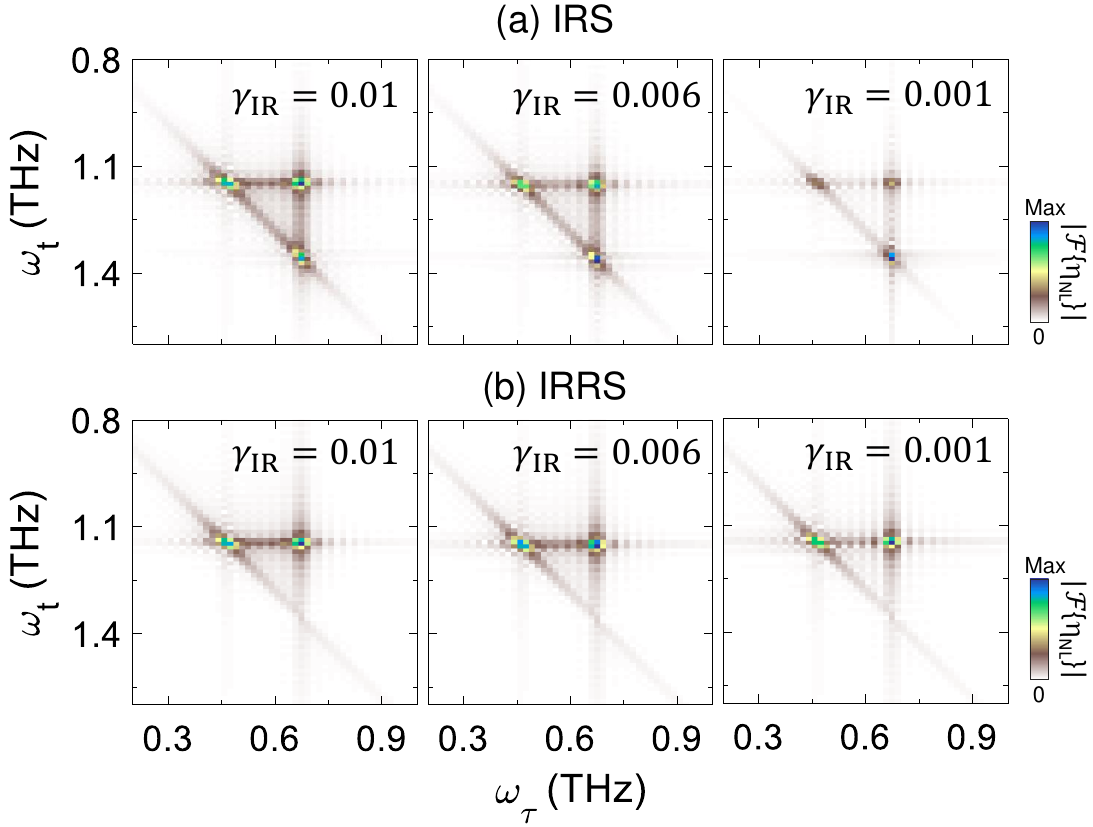}
    \caption{Dependence of the simulated spectra on the IR damping parameter $\gamma_{\rm IR}$. The top row shows the purely phononic driving of the Raman mode, $\propto Q_{\rm IR}^2$ (IRS), while the bottom row corresponds to the photo-phononic driving, $\propto Q_{\rm IR} E$ (IRRS). In each case, $\gamma_{\rm R} = 0.005$ is fixed, and $\gamma_{\rm IR}$ is varied.}
    \label{fig:app_fig1}
\end{figure}

We investigate the dependence of the simulated spectra on the damping of the IR-active phonon mode. Specifically, we compare the purely phononic driving term $Q_{\rm R} Q_{\rm IR}^2$ (IRS) and the photo–phononic driving term $Q_{\rm R} Q_{\rm IR} E$ (IRRS) for three representative damping constants ($\gamma_{\rm IR} = 0.001, 0.006,$ and $0.01$). The results are shown in Fig.~\ref{fig:app_fig1}.
As can be seen by comparing the peak intensities in the top row of Fig.~\ref{fig:app_fig1}, decreasing the IR damping $\gamma_{\rm IR}$ enhances the spectral feature at $\omega_t \sim 1.36~\text{THz} = 2\omega_{\rm IR}$ (peak B). This behavior arises because the IR mode oscillations persist longer at lower damping, and thereby more efficiently drive the Raman mode through the nonlinear force proportional to $Q_{\rm IR}^2$. In contrast, for larger $\gamma_{\rm IR}$, the IR mode predominantly follows the external electric field, resulting in a forced response $Q_{\rm IR}(t) \propto E(t)$. Consequently, the peaks at $\omega_t \sim 1.15~\text{THz}$ (corresponding to A1 and A2) become more pronounced. In this regime, the effective Raman driving term $\propto Q_{\rm IR}^2$ is reduced to a form proportional to $Q_{\rm IR} E$, favoring resonant excitation at the fundamental frequency.

The bottom row of Fig.~\ref{fig:app_fig1} shows the results for the photo-phononic IRRS mechanism, where the driving term is $Q_{\rm IR} E$. In this case, no competing energy transfer pathways are present, and the spectral profile remains essentially unchanged across different damping values. Notably, no additional high-frequency feature at $\omega_t \sim 1.36~\text{THz} = 2\omega_{\rm IR}$ emerges when the damping constant of the IR mode is reduced.

To better appreciate the competition between the two driving mechanisms at varying damping constants of the IR mode, we disentangle them analytically in App.~\ref{sec:app_analytics}.

\section{Analytical solutions to the equations of motion} \label{sec:app_analytics}


\subsection{IR-active mode driven by electric field}

The equation of motion for the displacement coordinate $Q$ of a linearly driven harmonic oscillator is $\ddot Q
+ 
\omega^2 Q
= 
F (t)$, where $\omega$ is the oscillator frequency and $F(t)$ is the driving term. Introducing the complex variable $\xi(t)
=
\dot Q (t)
+
i\omega Q (t)$ gives the general retarded solution 
\begin{equation}
Q (t)
=
\frac{1}{\omega}
\mathrm{Im}\,
\left\{
e^{i \omega t}
\int_{-\infty}^{t}
\mathrm{d} t' \,
F(t')
e^{-i \omega t'}
\right\}.
\label{eq:general_solution_for_driven_harmonic_oscillator}
\end{equation}
For the initial conditions, it was assumed that there are no pre-existing oscillations in the remote past so that $\xi (-\infty) = 0$, and that the motion is entirely driven by the force $F(t)$.

Neglecting the damping constant $\gamma_{\rm IR}$ and the back action of the coupling terms $\varpropto Q_{\rm IR} Q_{\rm R}$ and $\varpropto Q_{\rm R} E$, we get the driven harmonic oscillator equation
\begin{equation}
\ddot Q_{\rm IR}
+ 
\omega_{\rm IR}^2 Q_{\rm IR} 
= 
\frac{Z^*}{M_{\rm IR}} \, E(t),
\label{eq:eq_IR_mode}
\end{equation}
where $E(t)$ is the electric field of THz pump pulse. A good analytical approximation for the pump pulse is
\begin{equation}
    E (t) = E_0 \cos (\omega_{\rm F} t) e^{- t^2 / t_0^2}, \label{eq:analytical_appr_e_field}
\end{equation}
where $\omega_{\rm F}$ is the center frequency of THz field and $t_0$ is the pulse width. Because of the gaussian envelope, the electric field is localized in time. As a consequence, the integral for the general solution of the driven oscillator saturates for large times, $t \gg t_0$, and we can extend the upper time limit to infinity,

\begin{equation}
Q_{\rm IR}(t)
=
\frac{Z^*}{M_{\rm IR}\omega_{\rm IR}}
\mathrm{Im}
\left\{
e^{i\omega_{\rm IR}t}
\int_{-\infty}^{\infty}
\mathrm{d} t'\,
E(t')e^{-i\omega_{\rm IR}t'}
\right\}.
\label{eq:QIR_general_solution_to_infinite_limits}
\end{equation}
Rewriting $\cos (\omega_{\rm F} t) = (e^{i\omega_{\rm F} t} + e^{-i\omega_{\rm F} t})/2$ and using the general integral identity $
\int_{-\infty}^{\infty}
\mathrm{d} t \,
e^{-t^2/t_0^2}e^{-i\omega t}
=
\sqrt{\pi}t_0
\exp\left[
- t_0^2 \omega^2 / 4
\right]
$, we obtain that after a pulse, the driven infrared mode oscillates at its eigenfrequency, $Q_{\rm IR}(t)
=
A_{\rm IR}
\sin (\omega_{\rm IR} t )$. The oscillation amplitude $A_{\rm IR}
=
Z^* E_0 t_0\sqrt{\pi}
(
\exp (- t_0^2 (\omega_{\rm IR}-\omega_F)^2 / 4)
-
\exp (- t_0^2 (\omega_{\rm IR}+\omega_F)^2 / 4)
)
/ 
2M_{\rm IR}\omega_{\rm IR}$ is determined by the spectral overlap between the THz pulse and the phonon resonance. We note that the pulse may alternatively be approximated by a delta function, $E (t) = E_0 \cos (\omega_{\rm F} t) \delta (t)$, which yields $Q_{\rm IR} = Z^* E_0 \sin (\omega_{\rm IR} t) / ( M_{\rm IR}\omega_{\rm IR}) $. As expected, the resulting dynamics are qualitatively identical and differ only by a rescaling of the oscillation amplitude.

\subsection{Raman-active mode after two-pulse excitation of IR-active mode} \label{subsec:raman_after_two_pulse_excitation}

After two electric field pulses separated by a time $\tau$, the linear combination of the freely oscillating infrared active mode response is $Q_{\mathrm{IR}}(t,\tau)
=
A_{\mathrm{IR}}
(
\theta(t) \sin(\omega_{\mathrm{IR}} t)
+
\theta(t + \tau) \sin(\omega_{\mathrm{IR}} (t+\tau))
)$, assuming the convention that the movable pulse comes before the static one for $\tau > 0$. We note that the subsequent results remain invariant when switching all the signs for $\tau$, i.e. working with $\theta(t)\sin(\omega t)+\theta(t-\tau)\sin(\omega(t-\tau))$ and $\tau < 0$.
To obtain the driving term $F_{\rm R}(t,\tau)
=
Q_{\mathrm{IR}}^2(t,\tau)$ for the Raman mode in Eq.~\eqref{eq:equation_of_motion_Raman_mode}, we need to look at the squared IR mode response, 
\begin{equation}
\begin{split}
Q_{\mathrm{IR}}^2(t,\tau)
=
A_{\mathrm{IR}}^2
\Big[
&\theta (t) \sin^2(\omega_{\mathrm{IR}} t) \\
+&
\theta (t + \tau) \sin^2(\omega_{\mathrm{IR}} (t+\tau)) \\
+&
\theta (t) \theta (t + \tau) \cos(\omega_{\mathrm{IR}}\tau) \\
-&
\theta (t) \theta (t + \tau) \cos(2\omega_{\mathrm{IR}} t + \omega_{\mathrm{IR}}\tau)
\Big].
\end{split}
\end{equation}
This drive contains the indirect drives of the Raman mode from the coupling to the IR mode, coming from the two separate $A$ and $B$ pulses, as well as from both pulses. We are however only interested in the nonlinear Kerr signal of the Raman mode. Because adding up different drives is linear, we can define the nonlinear drive 
\begin{equation}
\begin{split}
F_{\rm R}^{\rm NL} (t,\tau)
= 
\frac{a A_{\rm IR}^2}{M_{\rm R}} 
\Big(
& \theta (t) \theta (t + \tau) \cos (\omega_{\rm IR}\tau) \\
-&
\theta (t) \theta (t + \tau) \cos (2\omega_{\mathrm{IR}} t
+
\omega_{\mathrm{IR}}\tau)
\Big).
\end{split}
\label{eq:nonlinear_drive_from_nonlinear_phononic_driving}
\end{equation}
Again neglecting the damping $\gamma_{\rm R}$ and integrating the general driven oscillator equation from Eq.~\eqref{eq:general_solution_for_driven_harmonic_oscillator} with the driving term in Eq.~\eqref{eq:nonlinear_drive_from_nonlinear_phononic_driving}, we obtain the nonlinear Raman signal
\begin{equation}
\begin{split}
Q_{\rm R}^{\rm NL}(t,\tau)
=& 
\frac{a A_{\mathrm{IR}}^2}{M_{\rm R}\omega_{\rm R}}
\Bigg\{
\frac{1}{\omega_{\rm R}}
\cos(\omega_{\rm IR}\tau)
\\
&-
\frac{2\omega_{\rm IR}}{2\omega_{\rm IR} \omega_{\rm R} - \omega_{\rm R}^2}
\cos(\omega_{\rm R}t+\omega_{\rm IR}\tau)
\\
&-
\frac{2\omega_{\rm IR}}{2\omega_{\rm IR} \omega_{\rm R} + \omega_{\rm R}^2}
\cos(\omega_{\rm R}t-\omega_{\rm IR}\tau)
\\
&+
\frac{\omega_{\rm R}}{4\omega_{\rm IR}^2-\omega_{\rm R}^2}
\cos\!\big(2\omega_{\rm IR}t+\omega_{\rm IR}\tau\big)
\Bigg\}.
\end{split}
\end{equation}
Identifying $\tau$ and $t$ as excitation and detection time, the two-dimensional Fourier transform gives peaks at
$(\omega_{\tau},\omega_{t}) = (\pm\omega_{\rm IR},\omega_{\rm R})$ and 
$(\omega_{\rm IR},2\omega_{\rm IR})$, which correspond to peaks A2 and B in Fig.~\ref{fig:3}(b). In particular, this calculation identifies peak B as coming from the purely phononic driving. 


\subsection{Raman-active mode dynamics under direct (via electric field) and indirect (via IR-active mode) driving} \label{subsec:raman_mode_after_field}

After excitation by the first pulse, the IR phonon oscillates as $
Q_{\mathrm{IR}}(t)=A_{\mathrm{IR}}\sin(\omega_{\mathrm{IR}}t)
$ with amplitude $A_{\mathrm{IR}}$ as derived before. However, the driving term $\varpropto Q_{\rm IR}^2$ is not in resonance with the Raman mode, $2\omega_{\rm IR} \neq \omega_{\rm R}$. Therefore,  we expect that during the overlap of the first and second pulse, the IRS driving term effectively reduces to the IRRS mechanism $\varpropto Q_{\rm IR} E$. That is, during the THz pulse excitation, the $Q_{\rm IR}$ normal coordinate undergoes a forced oscillation in which essentially $Q_{\rm IR} \varpropto E$. Therefore, while the THz pulse is present, one can effectively replace one $Q_{\rm IR}$ by $E$ in the anharmonic coupling term $Q_{\rm IR}^2$ and the resulting driving term $F_{\rm R} (t,\tau) = Q_{\rm IR} (t) E (t + \tau)$ fulfills the resonant condition due to the broad spectrum of the THz pulse. Considering a delta function pulse, $E (t) = E_0 \cos (\omega_{\rm F} t) \delta (t)$, and integrating the general driven oscillator equation from Eq.~\eqref{eq:general_solution_for_driven_harmonic_oscillator}, we obtain 

\begin{equation}
\begin{split}
Q_{\rm R} (t,\tau) =& 
-\frac{bA_{\mathrm{IR}}E_0}{2M_R\omega_R}
\\
\times \Big(&\cos (\omega_R t + (\omega_R-\omega_{\rm IR})\tau ) \\
-&
\cos (\omega_R t + (\omega_R+\omega_{\rm IR})\tau)
\Big).
\end{split}
\label{eq:raman_after_field_driving_1}
\end{equation}
We note that the peak positions in the frequency domains remain unchanged when using the more realistic pulse approximation of Eq.~\eqref{eq:analytical_appr_e_field}, which just affects the mode amplitudes. From the Raman response in Eq.~\eqref{eq:raman_after_field_driving_1}, we see that the Raman mode oscillates at its eigenfrequency $\omega_{\rm R}$, while the delay $\tau$ introduces phase modulations at frequencies $\omega_{\rm R} \pm \omega_{\rm IR}$. So far, this discussion assumed that the second THz pulse excites the Raman mode via the already excited IR mode of the first pulse through the nonlinear photo-phononic driving. On the other hand, we can consider the case where the THz pulse $E(t+\tau)$ arrives before the nominally first THz pulse $E(t)$. In this case, the driving term becomes $F_{\rm R} (t,\tau) = Q_{\rm IR} (t + \tau) E (t)$. Again solving the general driven oscillator equation from Eq.~\eqref{eq:general_solution_for_driven_harmonic_oscillator} with the delta function pulse for the field, we obtain 
\begin{equation}
\begin{split}
Q_{\rm R} (t,\tau) = 
\frac{bA_{\mathrm{IR}}E_0}{2M_R\omega_R} (\cos (\omega_R t -\omega_{\rm IR}\tau )
-
\cos (\omega_R t +\omega_{\rm IR}\tau)
),
\end{split}
\end{equation}
which is the same as Eq.~\eqref{eq:raman_after_field_driving_1} after substituting $t \to t - \tau$. We find that there are peaks at $(\omega_{\tau},\omega_{t}) = (\pm\omega_{\rm IR},\omega_{\rm R})$, where the positive branch corresponds to peak A2 in Fig.~\ref{fig:3}(b).

\section{Further considerations for quantum polarons}
\label{sec:foldy_wouthuysen_transformation}

\subsection{Background of Eq.~(\ref{eq:hamiltonian_spin_electric})}
The low-energy optoelectronic properties of lead-halide perovskites in the cubic phase follow from the hybridization of the $s$- and $p$-orbitals of lead and halide atoms~\cite{Umebayashi2003}. It is known that it is sufficient to consider only the following four basis states~\cite{Becker2018}: 
\begin{align}
|\DOWN\Uparrow\rangle&= |s\rangle |\uparrow\rangle, \label{eq:states_1}\\
|\DOWN\Downarrow\rangle& = |s\rangle |\downarrow\rangle \label{eq:states_2},\\
|\UP\Uparrow\rangle&=-\frac{|p_z\rangle |\uparrow\rangle+\left(|p_x\rangle + i|p_y\rangle\right) |\downarrow\rangle}{\sqrt{3}}, \label{eq:states_3}  \\ 
|\UP\Downarrow\rangle&= \frac{|p_z\rangle |\downarrow\rangle-\left(|p_x\rangle - i|p_y\rangle\right) |\uparrow\rangle}{\sqrt{3}}, \label{eq:states_4}
\end{align}  
where the left-hand-side introduces the `quasi-spin' notation convenient for our work. The right-hand-side follows the standard notation~\cite{Kane1966,Chuang1995} for the spin structure of a state [$|\downarrow\rangle$ and $|\uparrow\rangle$] and for the components of the Bloch functions [$|s\rangle$, $|p_x\rangle$, $|p_y\rangle$, $|p_z\rangle$]. 

The most general form of the Hamiltonian can be obtained by fixing time-reversal, and inversion symmetries of the system. It is straightforward to show that the operators that implement these symmetries for the states in Eqs.~(\ref{eq:states_1})-(\ref{eq:states_4}) are
\begin{align}
\hat T= i\tau_3\otimes \sigma_2 \hat K,\\
\hat P=-\tau_3\otimes\sigma_0,
\end{align}
where $\hat K$ is the complex conjugation operator, and the operators $\tau_i$ ($\sigma_i$ ) act on the $\DOWN/\UP$ ($\Downarrow/\Uparrow$) spaces. One can show in the presence of a cubic symmetry one cannot have a term in the Hamiltonian that is linear in $Q$ and electronic (spin and orbital) degrees of freedom. 

\subsection{Derivation of Eq.~(\ref{eq:deltaH})}
Here, we derive the correction to the effective Hamiltonian, $\delta H$ as written in Eq.~\eqref{eq:deltaH}. Near the band-gap point, we consider the Hamiltonian~\cite{Volosniev2023,Volosniev2023a} $H = H_0 + H_{\rm el-ph}$ with $H_0 = (\Delta_g  / 2) \tau_3\otimes\sigma_0$ and $H_{\rm el-ph}$ as in Eq.~\eqref{eq:deltaH}. We can remove the off-diagonal transitions across the band gap by constructing an effective low energy Hamiltonian $H_{\rm eff}$ \textit{via} the transformation $H_{\rm eff} = e^{iS} H e^{-iS}$, where the leading-order generator is chosen as $S = (\lambda_{\rm ep} / \Delta_g) \tau_2\otimes \vec{\sigma}\cdot\vec Q$. It can be verified that $i[S,H_0]=-H_{\rm el-ph}$ and consequently, applying the Baker-Campbell-Hausdorff formula gives
\begin{equation}
H_{\rm eff}
\simeq 
H_0+\frac{i}{2}[S,H_{\rm el-ph}],
\end{equation}
where the electron-phonon interaction term vanishes to first-order. The low energy correction to $H_0$ is therefore 
\begin{equation}
\delta H
\equiv
\frac{i}{2}[S,H_{\rm el-ph}]
=\frac{\lambda_{\rm ep}^2}{\Delta_g}
 \tau_3\otimes(\vec{\sigma}\cdot\vec Q)^2.
\end{equation}
Using the Pauli matrix identity $\sigma_i\sigma_j=\delta_{ij}\sigma_0 + i\sum_k\varepsilon_{ijk}\sigma_k$, and that antisymmetric combinations of the phonon coordinate vanish because $Q_i Q_j = Q_j Q_i$, we obtain the final result
\begin{equation}
\delta H
=\frac{\lambda_{\rm ep}^2Q_{\mathrm{IR}}^2}{\Delta_g}
 \tau_3\otimes\sigma_0,
\end{equation}
which corresponds to Eq.~\eqref{eq:deltaH} of the main text. This equation also follows from a Foldy-Wouthuysen transformation analogous to the one presented in Ref.~\cite{Kumar2025}. 
We note that in quantum optics, the transformation $e^{iS} H e^{-iS}$ is also known as Schrieffer-Wolff transformation~\cite{klimovEffectiveHamiltoniansQuantum2002}.

\bibliography{aipsamp}

\end{document}